\documentclass[%
 reprint,
 amsmath,amssymb,
 aps,
 prd,
 floats,
]{revtex4-1}

\usepackage{graphicx}% Include figure files
\usepackage{dcolumn}% Align table columns on decimal point
\usepackage{bm}% bold math
\begin{document}

\preprint{APS/123-QED}

\title{Anisotropic evolution of 4-brane in a 6d generalized Randall-Sundrum model}% Force line breaks with \\

\author{Guang-Zhen Kang}
\email{gzkang@nju.edu.cn}
\affiliation{School of Science, Yangzhou Polytechnic Institute, Yangzhou 225127, China}
\affiliation{Department of Physics, Nanjing University, Nanjing 210093, China}

\author{De-Sheng Zhang}
\affiliation{School of Science, Changzhou Institute of Technology, Changzhou 213032, China}

\author{Long Du}
\affiliation{School of Science, Guangxi University of Science and Technology, Liuzhou 545006, China}

\author{Jun Xu}
\affiliation{School of Science, Yangzhou Polytechnic Institute, Yangzhou 225127, China}

\author{Hong-Shi Zong}
\affiliation{Department of Physics, Nanjing University, Nanjing 210093, China}
\affiliation{Joint Center for Particle, Nuclear Physics and Cosmology, Nanjing 210093, China}
\affiliation{State Key Laboratory of Theoretical Physics, Institute of Theoretical Physics, CAS, Beijing 100190, China}

%\affiliation{
%\affiliation{
%\affiliation{%


\date{\today}% It is always \today, today,

\begin{abstract}
We investigate a 6d generalized Randall-Sundrum brane world scenario with a bulk cosmological constant.
It is shown that each stress-energy tensor $T_{ab}^{i}$ on the brane is similar to a constant vacuum energy.
This is consistent with the Randall-Sundrum model in which each 3-brane Lagrangian separated out a constant vacuum energy.
By adopting an anisotropic metric ansatz, we obtain the 5d Friedmann-Robertson-Walker field equations.
At a little later period, the expansion of the universe is proportional to $t^{\frac{1}{2}}$
which is as similar as the period of the radiation-dominated.
We also investigate the case with two $a(t)$ and two $b(t)$.
In a large region of $t$,
we obtain the 3d effective cosmological constant $\Lambda_{eff}=-2\Omega/3>0$ which is independent of the integral constant.
Here the scale factor is exponential expansion which is consistent with our present observation of the universe.
Our results demonstrate that it is possible to construct a model
which solves the dark energy problem, meanwhile guaranteeing a positive brane tension.
\begin{description}
\item[PACS numbers]
04.50.-h, 11.10.Kk, 11.25.Mj
\end{description}
\end{abstract}

\maketitle

%\tableofcontents

\section{Introduction}

In the early 1920s, Kaluza and Klein attempted to establish a more fundamental theory which unifies the forces of electromagnetism and gravitation by introducing extra dimension(s) into general relativity \cite{KK}.
The Kaluza-Klein (KK) theory attracted a lot of attention to
explore extra dimensions in various observable phenomena \cite{NAH1,NAH2,Antoniadis1,Antoniadis,Sundrum,Lykken}.
In the middle of last century, this interest in extra dimensions has been enhanced
because of the emergence of string/M theory in which the extra dimensional space appear naturally.
Inspired by the concept of brane in string theory \cite{Polchinski},
the braneworld scenario is proposed.
This theory can well explain some difficult problems in physics,
such as the hierarchy problem (the problem of why the electroweak scale/Higgs mass $M_{EW}\sim1$TeV is
so different from the Planck scale $M_{pl}\sim10^{16}$TeV) and the cosmological constant problem \cite{RS,ADD,Antoniadis,Das}.

The most successful resolution of hierarchy problem in the above theories is Randall-Sundrum (RS) two-brane model \cite{RS}.
The RS model takes into account the tension of the brane which causes the spacetime outside the brane to be curved.
It consists of a 5d AdS bulk with a negative cosmological constant $\Lambda$ and a single extra dimension
satisfying $S1/Z_{2}$ orbifold symmetry.
In such a scenario, our universe is described by a 5d metric
\begin{eqnarray}
ds^{2}=e^{-2\sigma(\phi)}\eta_{\mu\nu}dx^{\mu}dx^{\nu}+r_{c}^{2}d\phi^{2},
\end{eqnarray}
where $\phi$ is the coordinate for an extra dimension, $r_{c}$ is the compactification radius,
$e^{-2\sigma}$ is the warp factor with $\sigma=kr_{c}|\phi|$,
$k=\sqrt{-\Lambda/24M^{3}}$ with $M$ being the 5d Planck mass.
In this model, the weak scale is generated from the Planck scale through the warp factor which originates from the background metric.
But the visible brane in RS model have a negative tension which is intrinsically unstable.
Furthermore the visible 3-brane (four-dimensional spacetime) has zero cosmological constant,
which is not consistent with presently observed small value \cite{Das,Koley}.

Such a braneworld model has been widely studied.
It is shown that the induced cosmological constant and the brane tension of the visible brane can be both positive or negative \cite{Burgess,Nilles,Sasaki}.
By replacing $\eta_{\mu\nu}$ with $g_{\mu\nu}$, a generalized RS braneworld scenario is achieved \cite{Das}.
In this model, the negative brane cosmological constant is analysed in detail \cite{Mitra,SC1,SC2,SC3,Banerjee}.
It shows that $N$ has a minimum value $N_{min}=2n(n\approx16)$ which leads to an upper bound for the induced negative cosmological constants.
Furthermore, there are two different solutions to the hierarchical problem for a tiny value of cosmological constant.
One solution corresponds to the visible and hidden brane both with positive tension.
This is very interesting because both branes are stable.
In another case,
the induced positive cosmological constant corresponds to a negative tension visible brane which is instable,
so we do not considered this case anymore.

In above anti-de Sitter brane region,
a large part of the parameter space corresponds to a positive value for the visible brane tension.
But our universe is currently undergoing accelerated expansion which is indicated by recent observations of type Ia supernovae \cite{Perlmutter,Riess} and measurements of the anisotropies of the cosmic microwave background \cite{Bennett,Halverson,Netterfield}.
To explain this late-time epoch of accelerating expansion of the universe,
we assume that there is a cosmological constant component in 4d Einstein's field equation \cite{Middleton}.
The cosmological constant is a very small value ($\simeq10^{-124}$ in Planck unit) which is restricted by the above experiments.
So we need to cancel the induced negative cosmological constants in order to be consistent with observations.

In this paper,
we focus on a 6d braneworld models because there is no special reason to restrict the number of dimensions to five.
For solving the above problem of the induced cosmological constant on the visible brane being negative,
we consider the 4-brane (a extra dimension on the brane) in 6d generalized RS model instead of the 3-brane in 5d generalized RS model.
Then, we obtain the effective induced positive cosmological constant of 4d spacetime with an anisotropic metric ansatz.
At a little later period, the expansion of the 3d scale factor is as similar as the period of the radiation-dominated.
Our work is organized as follows: In Sec. \uppercase\expandafter{\romannumeral2},
by considering the 4-brane with the matter field Lagrangian in 6d generalized RS model,
we obtain a 5d Einstein field equation.
In Sec. \uppercase\expandafter{\romannumeral3},
we focus on the evolution of 4-brane solved from the above field equation with an anisotropic metric ansatz.
Finally, the summary and conclusion are presented in Sec. \uppercase\expandafter{\romannumeral4}.

\section{6d Generalized Randall-Sundrum Model}
We start with a 6d generalized Randall-Sundrum model action:
\begin{eqnarray}
S=S_{bulk}+S_{vis}+S_{hid}\label{eq:S}.
\end{eqnarray}
The bulk action, the visible brane action and the hidden brane action are respectively:
\begin{eqnarray}
S_{bulk}=\int d^{5}xdy\sqrt{-G}(M^{4}_{6}R-\Lambda)\label{eq:S1},\\
S_{vis}=\int d^{5}x\sqrt{-g_{vis}}(\mathcal{L}_{vis}-V_{vis}),\\
S_{hid}=\int d^{5}x\sqrt{-g_{hid}}(\mathcal{L}_{hid}-V_{hid}),
\label{eq:S2}
\end{eqnarray}
where $\Lambda$ is a bulk cosmological constant,
$M_{6}$ denotes 6d fundamental mass scale,
$G_{AB}$ is the 6d metric tensor,
$R$ is the 6d Ricci scalar,
$\mathcal{L}_{vis}(\mathcal{L}_{hid})$ and $V_{vis}(V_{hid})$ are the matter field Lagrangian
and the tension of the visible(hidden) brane, respectively.

Variation of the above action with respect to the 6d metric tensor $G_{AB}$ led to Einstein¡¯s equations:
\begin{eqnarray}\label{eq:field}
R_{AB}-\frac{1}{2}G_{AB}R=\frac{1}{2M^{4}_{6}}\{-G_{AB}\Lambda+\sum_{i}[T^{i}_{AB}\nonumber\\
\times\delta(y-y_{i})-G_{ab}\delta_{A}^{a}\delta_{B}^{b}V_{i}\delta(y-y_{i})]\},
\end{eqnarray}
where Capital Latin $A,B$ indices run over all spacetime dimensions, Lowercase Latin $a,b=0,1,2,3,4$
$R_{AB}$ and $T^{i}_{AB}$ are the 6d Ricci and the energy-momentum tensors respectively,
$y_{i}$ represents the position of the $i$-th brane in the sixth coordinate, $i=hid$ or $vis$.
The 6d stress-energy tensor $T^{iA}_{B}$ will be
assumed to be that of an anisotropic perfect fluid and of the form
\begin{eqnarray}
T^{iA}_{B}=diag[-\rho_{i}(t),p_{i1}(t),p_{i2}(t),p_{i3}(t),p_{i4}(t),0].
\end{eqnarray}

The metric ansatz in the generalized RS scenario,
satisfying the 6d Einstein equations is
\begin{eqnarray}\label{eq:ds2}
ds^{2}=G_{AB}dx^{A}dx^{B}=e^{-2A(y)}g_{ab}dx^{a}dx^{b}+r^{2}dy^{2},
\end{eqnarray}
where $g_{ab}$ is the 5d metric tensor.
The corresponding Einstein equations are given by:
\begin{eqnarray}\label{eq:field00}
\widetilde{R}=e^{-2A}(20A'^{2}+\frac{\Lambda}{M^{4}_{6}}),
\end{eqnarray}
and
\begin{eqnarray}\label{eq:fieldab}
\widetilde{R}_{ab}-\frac{1}{2}g_{ab}\widetilde{R}=g_{ab}e^{-2A}\{(4A''-10A'^{2})-\frac{1}{2M^{4}_{6}}[\Lambda\nonumber\\
+\sum_{i}\delta(y-y_{i})V_{i}]\}+\frac{e^{-2A}}{M^{4}_{6}}\sum_{i}T_{ab}^{i}\delta(y-y_{i}),
\end{eqnarray}
where $\widetilde{R}_{ab}$ and $\widetilde{R}$ are the five-dimensional Ricci tensor
and Ricci scalar respectively, defined with respect to $g_{\mu\nu}$.
One side of Eq.~(\ref{eq:field00}) contains the derivatives of $A(y)$,
depending on the extra coordinate $y$ alone,
while the other side depends on the brane coordinates $x_{\mu}$ alone.
Thus each side is equal to an arbitrary constant.
For convenience, we take this arbitrary constant to be $10\Omega/3$.
Thus, we get from Eq.~(\ref{eq:field00}):
\begin{eqnarray}\label{eq:field001}
e^{-2A}(20A'^{2}+\frac{\Lambda}{M^{4}_{6}})=\frac{10}{3}\Omega,
\end{eqnarray}
and
\begin{eqnarray}\label{eq:field002}
\widetilde{R}=\frac{10}{3}\Omega.
\end{eqnarray}

Multiply both sides of Eq.~(\ref{eq:fieldab}) by $g^{ab}$,
and rearranging terms, we get:
\begin{eqnarray}\label{eq:fieldab1}
\widetilde{R}=-\frac{2}{3}e^{-2A}\{(4A''-10A'^{2})-\frac{1}{2M^{4}_{6}}[\Lambda\nonumber\\
+\sum_{i}\delta(y-y_{i})V_{i}]\}-\frac{e^{-2A}}{3M^{4}_{6}}\sum_{i}T^{i}\delta(y-y_{i}),
\end{eqnarray}
where $T^{i}=g^{ab}T^{i}_{ab}$.
Using Eqs.~(\ref{eq:field001}) and (\ref{eq:field002}) cancel $A'^{2}$ and $\widetilde{R}$ in Eq.~(\ref{eq:fieldab1}),
yielding a simplified expression for $A''$,
\begin{eqnarray}\label{eq:fieldab2}
A''=\frac{\Omega}{6}e^{2A}+\frac{1}{8M^{4}_{6}}\sum_{i}\delta(y-y_{i})(V_{i}-\frac{T^{i}}{5}).
\end{eqnarray}
The left side and the first term of the right depending on the extra coordinate $y$ alone,
while the other term appeare only when the extra coordinate $y=y_{i}$ alone.
Thus we get $T^{i}=constant$. For convenience, we define $T^{i}\equiv5C_{i}$, where the the $C_{i}$ is a constant.
Eq.~(\ref{eq:fieldab2}) can be written:
\begin{equation}\label{eq:field123}
A''=\dfrac{\Omega}{6}e^{2A}+\frac{1}{8M^{4}_{6}}\sum_{i}\delta(y-y_{i})(V_{i}-C_{i}).
\end{equation}
Rearrange Eq.~(\ref{eq:field001}), we get an expression for $A'^{2}$:
\begin{equation}\label{eq:field1234}
A'^{2}=\dfrac{\Omega}{6}e^{2A}+k^{2},
\end{equation}
where $k^{2}\equiv-\Lambda/20M^{4}_{6}>0$ (for Ads bulk i.e. $\Lambda<0$).
We cancel the $A'^{2}$ and $A''$ in eq.~(\ref{eq:fieldab}) by Eqs.~(\ref{eq:field123}) and (\ref{eq:field1234}),
then get:
\begin{eqnarray}\label{eq:fieldabb}
\widetilde{R}_{ab}-\frac{1}{2}g_{ab}\widetilde{R}=-\Omega g_{ab}+\frac{1}{2M^{4}_{6}}\nonumber\\
\times\sum_{i}(T_{ab}^{i}-g_{ab}C_{i})\delta(y-y_{i}).
\end{eqnarray}
From the above equation, we can see that $T_{ab}^{i}-g_{ab}C_{i}=0$,
then we get $\rho_{i}=-p_{i1}=-p_{i2}=-p_{i3}=-p_{i4}=-C_{i}$.
So each stress-energy tensor $T_{ab}^{i}$ is similar to a constant vacuum energy.
This is consistent with the RS model \cite{RS} in which each 3-brane Lagrangian separated out a constant vacuum energy.
We define the $\mathcal{V}_{i}\equiv V_{i}-C_{i}$.
Thus, we get a 5d Einstein field equation:
\begin{equation}\label{eq:5dfield}
\widetilde{R}_{ab}-\frac{1}{2}g_{ab}\widetilde{R}=-\Omega g_{ab},
\end{equation}
and the system of equations of $A(y)''$ and $A'^{2}$:
\begin{equation}
\left\{
\begin{aligned}\label{eq:field122}
A''&=\dfrac{\Omega}{6}e^{2A}+\frac{1}{8M^{4}_{6}}\sum_{i}\delta(y-y_{i})\mathcal{V}_{i}, \\
A'^{2}&=\dfrac{\Omega}{6}e^{2A}+k^{2}. \\
\end{aligned}
\right.
\end{equation}
The above corresponds to the induced cosmological constant $\Omega$ on the visible brane.
For the induced brane cosmological constant $\Omega>0$ and $\Omega<0$,
the brane metric $g_{ab}$ may correspond to dS-Schwarzschild and AdS-Schwarzschild spacetimes respectively \cite{Karch}.
We first consider the induced negative cosmological constant on the visible brane,
the following solution for the warp factor is obtained:
\begin{eqnarray}\label{eq:solution-}
A=-\ln[\omega\cosh(k|y|+C_{-})],
\end{eqnarray}
where $\omega\equiv-\Omega/6k^{2}$,
and the constant $C_{-}=\ln\frac{1-\sqrt{1-\omega^{2}}}{\omega}$ for
considering the normalization of this factor at the orbifold fixed point $y = 0$.
Note in the limit $\omega\sim0$, the RS solution $A = ky$ can be recovered.
This is consistent with the results in Ref.~\cite{Das}.
The other solution $C_{-}=\ln\frac{1+\sqrt{1-\omega^{2}}}{\omega}$ is excluded
because the RS solution can not be recovered in the $\omega^{2}\rightarrow0$ limit.

We can get the 5d effective theory from the original action Eq.~(\ref{eq:S1}).
We focus on the curvature term from which we can derive the scale of gravitational interactions:
\begin{eqnarray}\label{eq:Seff}
S_{eff}\supset\int d^{5}x\int_{-\pi}^{\pi}dy\sqrt{-g}M^{4}_{6}re^{-3A(kry)}\tilde{R},
\end{eqnarray}
where we only focus on the coefficient proportional to five-dimensional Ricci scalar $\tilde{R}$.
The Legendre term \cite{Wu} is not proportional to $\tilde{R}$ when the metric was substituted inside the action.
So we do not consider this term here.
We can perform the $y$ integral to obtain a 5d action.
From this, we get
\begin{eqnarray}\label{eq:Seff1}
M^{3}_{5pl}=M^{4}_{6}[\frac{\omega^{6}}{12kc_{1}^{3}}(e^{3kr\pi}-1)+\frac{c_{1}^{3}}{12k}(1-e^{-3kr\pi})\nonumber\\
+\frac{3\omega^{4}}{4kc_{1}}(e^{kr\pi}-1)+\frac{3\omega^{2}c_{1}}{4k}(1-e^{-kr\pi})],
\end{eqnarray}
where $c_{1}\equiv1+\sqrt{1-\omega^{2}}$.
We find that if $\omega^{6}\ll e^{-3kr\pi}$, then $M_{5pl}$ depends only weakly on $r$ in the large $kr$ limit.
From this, Eq.~(\ref{eq:Seff1}) can be simplified to
\begin{eqnarray}\label{M3}
M^{3}_{5pl}=\frac{2M^{4}_{6}}{3k}(1-e^{-3kr\pi}).
\end{eqnarray}
Then we can get $M^{3}\simeq2M^{4}_{6}/3k$ in the large $kr$ limit.
In this 4-brane model, note the 5d components of the bulk metric is $g^{vis}_{ab}=G_{\mu\nu}(x^{a},y=r\pi)$,
we obtain:
\begin{eqnarray}\label{g5}
g^{vis}_{ab}=g_{ab}e^{-2A(kr\pi)},   \\
\sqrt{-g_{vis}}=\sqrt{-g}e^{-3A(kr\pi)}.
\end{eqnarray}
From the above equations, we can not determine the physical masses by properly normalizing the fields,
namely the hierarchy problem cannot be solved in this 4-brane model.

Take the second derivative of Eq.~(\ref{eq:solution-}) with respect to $y$, we get:
\begin{eqnarray}\label{eq:fieldAAA}
A''=\frac{\Omega}{6}e^{2A}-2k\tanh(k|y|+\ln\frac{1-\sqrt{1-\omega^{2}}}{\omega})\nonumber\\
\times(\delta(y)-\delta(y-y_{vis})).
\end{eqnarray}
Note the orbifold fixed point $y_{hid}=0$.
Comparing the above equation with Eq.~(\ref{eq:field122}),
we get the tension of the visible(hidden) $V_{vis}$ ($V_{hid}$):
\begin{eqnarray}\label{eq:Vvis}
V_{vis}=16M^{4}_{6}k\big[\frac{e^{2kr\pi}\frac{\omega^{2}}{c_{1}^{2}}-1}
{e^{2kr\pi}\frac{\omega^{2}}{c_{1}^{2}}+1}\big],
\end{eqnarray}
and
\begin{eqnarray}\label{eq:Vhid}
V_{hid}=16M^{4}_{6}k\big[\frac{1-\frac{\omega^{2}}{c_{1}^{2}}}{1+\frac{\omega^{2}}{c_{1}^{2}}}\big].
\end{eqnarray}
Setting $e^{-A(r\pi)}=10^{-n}$,
then we get from Eq.~(\ref{eq:solution-}):
\begin{eqnarray}\label{eq:n3}
10^{-N}=4(10^{-n}e^{-x}-e^{-2x}),
\end{eqnarray}
\begin{eqnarray}\label{eq:x2}
e^{-x}=\frac{10^{-n}}{2}[1\pm\sqrt{1-10^{-(N-2n)}}],
\end{eqnarray}
where $x\equiv\pi kr$, $\omega^{2}\equiv10^{-N}$.
For $1-10^{2n}\omega^{2}\geq0$,
we find $\omega^{2}\leq10^{2n}$ which leads to an upper bound for the cosmological constant ($\sim10^{-N}$) given by $N_{min}=2n$.
Eq.~(\ref{eq:x2}) have two values of $x$ instead of one, the both values give rise to the required warping.
For $(N-2n)\gg1$, the first solution of $x$ corresponds to the RS value plus a minute correction
which is given by $x_{1}=n\ln10+\frac{1}{4}10^{-(N-2n)}$,
while the second solution of $x$ is given by $x_{2}=(N-n)\ln10+\ln4$ \cite{Das}.
Obviously, the $x_{2}$ is greater than the $x_{1}$.
Rewriting Eq.~(\ref{eq:Vvis}) with $n$ and $N$, we get:
\begin{eqnarray}\label{eq:Vvis1}
\mathcal{V}_{vis}=16M^{4}_{6}k\frac{1-10^{N-2n}[1\pm\sqrt{1-10^{-(N-2n)}}]}{10^{N-2n}[1\pm\sqrt{1-10^{-(N-2n)}}]},
\end{eqnarray}
where the visible brane tension $\mathcal{V}_{vis}$ is different from Eq. (23) worked out in Ref.~\cite{Das}.
The two brane tensions are approximately given as:
\begin{eqnarray}\label{eq:Vvis-1}
\mathcal{V}_{vis-1}=-16M^{4}_{6}k,
\end{eqnarray}
\begin{eqnarray}\label{eq:Vvis-2}
\mathcal{V}_{vis-2}=16M^{4}_{6}k.
\end{eqnarray}
The visible brane tension in Eq.~(\ref{eq:Vvis-2}) is greater than Eq. (23) in Ref.~\cite{Das}
because that the denominator of Eq.~(\ref{eq:Vvis1}) is different from that of Eq. (23) in Ref.~\cite{Das}.
We see that a small negative cosmological constant suffices to render the tension positive,
provided the distance between the branes is somewhat larger than the value predicted by RS model.
Note the tension $\mathcal{V}_{vis-2}$ on the visible brane  is inconsistent with Eq. (25) in Ref.~\cite{Das}.
Because of $\omega\equiv10^{-N}\ll0$, we get that the hidden brane tension $V_{hid}$ is always positive.

For $\Omega>0$, the warp factor which satisfies Eq.~(\ref{eq:field122}) is given by:
\begin{eqnarray}\label{eq:solution+}
A=-\ln[\omega\sinh(-k|y|+C_{+})],
\end{eqnarray}
where $\omega\equiv\Omega/6k^{2}$, $C_{+}=\ln\frac{1+\sqrt{1+\omega^{2}}}{\omega}$.
In this case, the value of $\omega^{2}$ is unbounded,
so the positive brane cosmological constant $\Omega$ can be of arbitrary value.
The solution of $kr\pi$ is a single solution instead of two solutions for $\Omega<0$.
And the above solution is depend on $\omega^{2}$ and $n$.
For $\Omega>0$, the visible brane tension $\mathcal{V}_{vis}$ and the hidden brane tension are always negative and positive respectively \cite{Das}.
The negative tension visible brane is instable, so we do not considered this case anymore.

\section{Anisotropic evolution of 4-brane}
For $\Omega<0$, interestingly one can obtain the upper bound ($\sim-10^{-2n}$ in Planck units) of the induced negative cosmological constant on the visible 4-brane and the 4-brane tension can be positive for the second solution.
In this paper, we only consider two different spatial scaling factors $a(t)$ and $b(t)$.
\subsection{Case \uppercase\expandafter{\romannumeral1}}
First, we investigate the case with three $a(t)$ and one $b(t)$ which is most in line with the presently observed 3d space universe.
We choose an anisotropic metric ansatz of the form $g_{ab}=diag[-1,a^{2}(t),a^{2}(t),a^{2}(t),b^{2}(t)]$ \cite{Middleton}.
We allow the scale factor of the extra dimension on the visible brane $b(t)$ to evolve at a rate different from
that of the 3d scale factor $a(t)$.
This metric describes a flat, homogeneous, and isotropic 3d space and a flat extra dimension on the visible brane.
In this case, by adopting the above metric ansatz, we obtain the 5d FRW field equations from the Einstein field equations Eq.~(\ref{eq:5dfield}):
\begin{eqnarray}\label{eq:5dfield00}
H_{a}^{2}+H_{a}H_{b}=\frac{1}{3}\Omega,
\end{eqnarray}
\begin{eqnarray}\label{eq:5dfield44}
\dot{H}_{a}+2H_{a}^{2}=\frac{1}{3}\Omega,
\end{eqnarray}
\begin{eqnarray}\label{eq:5dfieldij}
2\dot{H}_{a}+\dot{H}_{b}+3H_{a}^{2}+H_{b}^{2}+2H_{a}H_{b}=\Omega,
\end{eqnarray}
where a dot denotes a time derivative,
$H_{a}\equiv\dot{a}/a$ and $H_{b}\equiv\dot{b}/b$ are the Hubble parameters of the 3d space and extra dimension respectively.
Eq.~(\ref{eq:5dfield44}) can be rewritten as:
\begin{eqnarray}\label{eq:5dfield442}
\frac{dH_{a}}{H_{a}^{2}-\frac{1}{6}\Omega}=-2dt.
\end{eqnarray}
Eq.~(\ref{eq:5dfield44}) can be integrated, then we get the following solution for the 3d Hubble parameter:
\begin{eqnarray}\label{eq:5dHa}
H_{a}=-\sqrt{-\frac{\Omega}{6}}\tan(2\sqrt{-\frac{\Omega}{6}}t+c),
\end{eqnarray}
where $c$ is a arbitrary constants of integration.
Performing the integration of Eq.~(\ref{eq:5dHa}), one find the solution of 3d space scale factor $a(t)$:
\begin{eqnarray}\label{eq:5da}
a=c_{a}\big|\cos(2\sqrt{-\frac{\Omega}{6}}t+c)\big|^{\frac{1}{2}},
\end{eqnarray}
where $c_{a}$ is a arbitrary constants of integration also.
We set that at the initial time $t=0$, $a=a_{0}$.
We can get $c_{a}=a_{0}|\cos c|^{-\frac{1}{2}}$,
Eq.~(\ref{eq:5da}) may then be rewritten as:
\begin{eqnarray}\label{eq:5daa}
a=a_{0}\frac{|\cos(2\sqrt{-\frac{\Omega}{6}}t+c)|^{\frac{1}{2}}}{|\cos c|^{\frac{1}{2}}}.
\end{eqnarray}
where the scale factor $a(t)$ increases with the increasing of $t$ when $-\frac{\pi}{2}<2\sqrt{-\frac{\Omega}{6}}t+c<0$.
For $\Omega<0$, the induced negative cosmological constant is bounded from below by $\sim-10^{-2n}$.
In order that the 3d space factor changes with time as smooth as possible,
we get the second solution $x_{2}\simeq(N-n)\ln 10+\ln 4\simeq172$ with $n\simeq50$ and $N\simeq124$.
Note here the above case $n\simeq50$ and $N\simeq124$ is satisfied both conditions $N-n\ll0$ from $\omega^{6}\ll e^{-3kr\pi}$ and $(N-2n)\gg1$.
In this case of $\Omega\simeq-10^{-124}$, we obtain that $2\sqrt{-\frac{\Omega}{6}}t\ll1$
when $t$ is not very large (for today $t\sim10^{60}$ in Planck unit).
We set the constants $c$ in Eq.~(\ref{eq:5daa}) equal to $-\pi/2$ plus a small positive constant $\chi$ to make sure that
the scale factor $a(t)$ is increasing from $t=0$ to the present $t\sim10^{60}$.
Then the scale factor $a(t)$ can be written:
\begin{eqnarray}\label{eq:earlya}
a&=&a_{0}\frac{|\cos[2\sqrt{-\frac{\Omega}{6}}t-\frac{\pi}{2}+\chi]|^{\frac{1}{2}}}{|\cos (-\frac{\pi}{2}+\chi)|^{\frac{1}{2}}}\nonumber\\
&=&a_{0}\frac{\sin^{\frac{1}{2}}(2\sqrt{-\frac{\Omega}{6}}t+\chi)}{\sin^{\frac{1}{2}}\chi}.
\end{eqnarray}
The Hubble parameter $H_{a}$ is rewritten as:
\begin{eqnarray}\label{eq:earlyHa}
H_{a}&=&-\sqrt{-\frac{\Omega}{6}}\tan(2\sqrt{-\frac{\Omega}{6}}t-\frac{\pi}{2}+\chi)\nonumber\\
&=&\sqrt{-\frac{\Omega}{6}}\cot(2\sqrt{-\frac{\Omega}{6}}t+\chi).
\end{eqnarray}
When $\sqrt{-\frac{\Omega}{6}}t\ll\chi\ll1$,
the Hubble parameter $H_{a}$ is obtained:
\begin{eqnarray}\label{eq:earlyHa2}
H_{a}\simeq\sqrt{-\frac{\Omega}{6}}\frac{1}{\chi}.
\end{eqnarray}
Here $H_{a}$ is a constant leading to an exponential expansion of the 3d scale factor.
Comparing with the FRW equation in which the gaussian curvature $K=0$, and contains only the cosmological constant,
we obtain the 4d effective cosmological constant $\Lambda_{eff}=-\Omega/2\chi^{2}>0$.
But the above period is so short that the 3d space scale factor only increase from $a_{0}$ to $a_{0}(1+\sqrt{\frac{-\Omega}{6\chi^{2}}}t)$.
After that, we obtain the Hubble parameter $H_{a}$ and the 3d scale factor $a(t)$ when $\chi\ll\sqrt{-\frac{\Omega}{6}}t\ll1$:
\begin{eqnarray}
H_{a}&\simeq&\frac{1}{2t}\label{eq:earlyHa3},\\
a(t)&\simeq&a_{0}(-\frac{2\Omega}{3\chi^2})^{\frac{1}{4}}t^{\frac{1}{2}}\label{eq:earlya3},
\end{eqnarray}
where $a(t)$ is proportional to $t^{\frac{1}{2}}$
which is as similar as the period of the radiation-dominated.
The deceleration parameter $q\equiv-\ddot{a}a/\dot{a}^{2}=1-\Omega/3H_{a}^{2}>1$.
It is unsatisfactory because the aforementioned deceleration parameter is not consistent with currently undergoing accelerated expansion.

Using Eqs.~(\ref{eq:5dfield00}) and ~(\ref{eq:earlyHa}),
the extra dimension Hubble parameter $H_{b}$ is given by:
\begin{eqnarray}\label{eq:5dfieldhbha}
H_{b}&=&\frac{\Omega}{3H_{a}}-H_{a}\nonumber\\
&=&\frac{\Omega}{3\sqrt{-\frac{\Omega}{6}}\cot(2\sqrt{-\frac{\Omega}{6}}t+\chi)}\nonumber\\
&&-\sqrt{-\frac{\Omega}{6}}\cot(2\sqrt{-\frac{\Omega}{6}}t+\chi).
\end{eqnarray}
Note when $H_{a}>0$, we obtain $H_{b}<0$, and vice versa.
Performing the integration of Eq.~(\ref{eq:5dfieldhbha}), one find the solution of the extra dimension scale factor $b(t)$:
\begin{eqnarray}\label{eq:5dbb}
b=b_{0}\frac{\sin^{\frac{1}{2}}\chi\cos(2\sqrt{-\frac{\Omega}{6}}t+\chi)}{\cos\chi\sin^{\frac{1}{2}}(2\sqrt{-\frac{\Omega}{6}}t+\chi)},
\end{eqnarray}
where we considered the initial conditions that when time $t=0$, $b=b_{0}$.
Form Eqs.~(\ref{eq:5daa}) and (\ref{eq:5dbb}),
it is obvious that the scale factor $a(t)$ and $b(t)$ are impossible to increase or reduce at the same time.
When the scale factor $a(t)$ increases, $b(t)$ decreases, and vice versa.
In other words, the decrease of $b(t)$ provides a driving force for the increasing of $a(t)$.

The above investigation is the case of the increasing of the 3d scale factor $a(t)$.
In the following, we investigate the case the scale factor $a(t)$ decrease with the increasing of time $t$ when $0<2\sqrt{-\frac{\Omega}{6}}t+c<\frac{\pi}{2}$ in Eq.~(\ref{eq:5daa}).
The analysis is similar to the previous one,
we substitute a small positive constant $\psi$ into the constants $c$ in Eq.~(\ref{eq:5daa}).
The Hubble parameter $H_{a}$ and $H_{b}$ are rewritten as:
\begin{eqnarray}\label{eq:earlyHa1}
H_{a}=-\sqrt{-\frac{\Omega}{6}}\tan(2\sqrt{-\frac{\Omega}{6}}t+\psi),
\end{eqnarray}
\begin{eqnarray}\label{eq:5dfieldhbha1}
H_{b}&=&\frac{\Omega}{3H_{a}}-H_{a}\nonumber\\
&=&-\frac{\Omega}{\sqrt{-\frac{\Omega}{6}}\tan(2\sqrt{-\frac{\Omega}{6}}t+\psi)}\nonumber\\
&&-\sqrt{-\frac{\Omega}{6}}\tan(2\sqrt{-\frac{\Omega}{6}}t+\psi).
\end{eqnarray}
Then we obtain the scale factors $a(t)$ and $b(t)$:
\begin{eqnarray}\label{eq:earlya1}
a=a_{0}\frac{\cos^{\frac{1}{2}}(2\sqrt{-\frac{\Omega}{6}}t+\psi)}{\cos^{\frac{1}{2}}\psi},
\end{eqnarray}
\begin{eqnarray}\label{eq:5dbb1}
b=b_{0}\frac{\cos^{\frac{1}{2}}\psi\sin(2\sqrt{-\frac{\Omega}{6}}t+\psi)}{\sin\psi\cos^{\frac{1}{2}}(2\sqrt{-\frac{\Omega}{6}}t+\psi)}.
\end{eqnarray}
When $\sqrt{-\frac{\Omega}{6}}t\ll\psi\ll1$,
the Hubble parameter $H_{a}\simeq-\sqrt{-\frac{\Omega}{6}}\psi$ is a negative constant.
The Hubble parameter $H_{b}$ is obtained:
\begin{eqnarray}\label{eq:earlyhb2}
H_{b}\simeq2\sqrt{-\frac{\Omega}{6}}\frac{1}{\psi}.
\end{eqnarray}
Then we obtain the Hubble parameter $H_{b}$ and $b(t)$ when $\psi\ll\sqrt{-\frac{\Omega}{6}}t\ll1$:
\begin{eqnarray}
&&H_{b}\simeq\frac{1}{t}\label{eq:earlyHb3},\\
&&b\simeq b_{0}(-\frac{2\Omega}{3\psi^2})^{\frac{1}{2}}t\label{eq:earlyb3},
\end{eqnarray}
where $b(t)$ is proportional to $t$ which is faster than the $a(t)$ in the case of the increasing of $a(t)$.
Because the decreasing of three dimensions instead of one provides dynamic.
In Case \uppercase\expandafter{\romannumeral1}, the decreasing of scale factor(s) on the brane does(do) not provide sufficient impetus for the other scale factors(factor) to  expand exponentially.

\subsection{Case \uppercase\expandafter{\romannumeral2}}

Similarly to the Case \uppercase\expandafter{\romannumeral1},
we investigate the case with two $a(t)$ and two $b(t)$.
We choose an anisotropic metric ansatz of the form $g_{ab}=diag[-1,a^{2}(t),a^{2}(t),b^{2}(t),b^{2}(t)]$,
the 5d FRW field equations are of the form:
\begin{eqnarray}\label{eq:5dfield00aabb}
H_{a}^{2}+4H_{a}H_{b}+H_{b}^{2}=\Omega,
\end{eqnarray}
\begin{eqnarray}\label{eq:5dfield44aabb}
\dot{H}_{a}+2\dot{H}_{b}+H_{a}^{2}+3H_{b}^{2}+2H_{a}H_{b}=\Omega,
\end{eqnarray}
\begin{eqnarray}\label{eq:5dfieldijaabb}
\dot{H}_{b}+2\dot{H}_{a}+H_{b}^{2}+3H_{a}^{2}+2H_{a}H_{b}=\Omega.
\end{eqnarray}
where the $H_{a}$ and $H_{b}$ are symmetric.
Setting $H_{a}$ positive and $H_{b}$ negative, we obtain the following solutions for the Hubble parameters $H_{a}$ and $H_{b}$ respectively:
\begin{eqnarray}
H_{a}&=&-\sqrt{\frac{-\Omega}{6}}[\tan(2\sqrt{\frac{-2\Omega}{3}}t+c_{2})\nonumber\\
&&-\sqrt{3}|\sec(2\sqrt{\frac{-2\Omega}{3}}t+c_{2})|]\label{eq:5dfieldHa22},\\
H_{b}&=&-\sqrt{\frac{-\Omega}{6}}[\tan(2\sqrt{\frac{-2\Omega}{3}}t+c_{2})\nonumber\\
&&+\sqrt{3}|\sec(2\sqrt{\frac{-2\Omega}{3}}t+c_{2})|]\label{eq:5dfieldHb22}.
\end{eqnarray}

\begin{figure}
\includegraphics[scale=0.5]{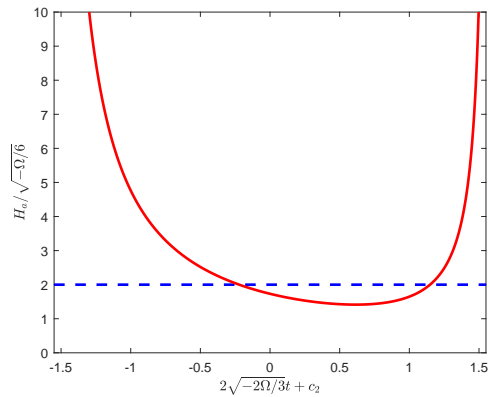}
\caption{\label{fig:1} (Color online). The Hubble parameter $H_{a}$ (solid curve) varies as a function of $2\sqrt{-2\Omega/3}t+c_{2}$ in Case \uppercase\expandafter{\romannumeral2}. The dashed curve is a constant $H\simeq2$.}
\end{figure}

As show in Fig.~\ref{fig:1}, the Hubble parameter $H_{a}$ is close to a constant $H\simeq2$ in a large region $-0.8<2\sqrt{-2\Omega/3}t+c_{2}<1.2$.
It is very different from the Case \uppercase\expandafter{\romannumeral1} in which the Hubble parameter $H_{a}$ is a constant in a very tiny interval $\sqrt{-\frac{\Omega}{6}}t\ll\chi\ll1$.
Considering the constraints in the Eqs.~(\ref{eq:5dfieldHa22}) and (\ref{eq:5dfieldHb22}) as in Case \uppercase\expandafter{\romannumeral1},
$2\sqrt{-\frac{2\Omega}{3}}t$ changes slowly with $t$ when we also set the second solution $x_{2}\simeq(N-n)\ln 10+\ln 4\simeq172$ with $n\simeq50$ and $N\simeq124$.
We obtain the 3d effective cosmological constant $\Lambda_{eff}=-2\Omega/3>0$ which is independent of the integral constant.
This is an important result.
It tells us that we can obtain an exponential expansion solution which is consistent with our presently observed universe when we start from a induced negative cosmological constants on the brane.
It is unsatisfactory because that the numbers of the expansion scale factor is two.
But this problem should be solved in a higher dimensional brane.

Finally, we consider an isotropic metric ansatz of the form
$g_{ab}=diag[-1,a^{2}(t),a^{2}(t),a^{2}(t),a^{2}(t)]$ in the 5d Einstein field equations Eq.~(\ref{eq:5dfield}),
then we obtain the time-time component of 5d FRW field equations:
\begin{eqnarray}\label{eq:5dfield00aa}
H_{a}^{2}=\frac{1}{6}\Omega.
\end{eqnarray}
Note there is no solution to the above equation because $\Omega<0$.

\section{Summary and Conclusion}

To summarize, in this paper we investigate a 6d theory with a 4-brane in order to solve the cosmological fine tuning problem.
We find that each stress-energy tensor $T_{ab}^{i}$ on the brane is similar to a constant vacuum energy.
Note the hierarchy problem cannot be solved well in this model.
This is consistent with the RS model \cite{RS} in which each 3-brane Lagrangian separates out a constant vacuum energy.
The visible brane tension obtained in our paper is greater than the result in Ref.~\cite{Das}.
For $\Omega<0$, the induced negative cosmological constant on the visible 4-brane has an upper bound ($\sim-10^{-32}$ in Planck units),
and the 4-brane tension is positive for the second solution.

In above case, we obtain the 5d FRW field equations from the Einstein field equations by adopting an anisotropic metric ansatz.
In Case \uppercase\expandafter{\romannumeral1}, we find that the 3d space scale factor is increasing from $t=0$ to the present $t\sim10^{60}$.
The constant Hubble parameter resulted in exponential expansion of the 3d scale factor slightly after the initial time $t=0$.
But the period is so short that 3d space scale factor only increases from $a_{0}$ to $a_{0}(1+\sqrt{\frac{-\Omega}{6\chi^{2}}}t)$.
When $\chi\ll\sqrt{-\frac{\Omega}{6}}t\ll1$, 3d space scale factor $a(t)$ is proportional to $t^{\frac{1}{2}}$
which is as similar as the period of the radiation-dominated.

In Case \uppercase\expandafter{\romannumeral2},
we investigate the case with two $a(t)$ and two $b(t)$.
In a large region of $t$,
we obtain the 3d effective cosmological constant $\Lambda_{eff}=-2\Omega/3>0$ which is independent of the integral constant.
Here the scale factor is of exponential expansion which is consistent with our presently observed universe.
It is shown that the expansion rate of scale factor is not directly related to the numbers of the scale factor of decrease.
It is unsatisfactory because that the numbers of the expansion scale factor is two.
But this problem should be solved in a higher dimensional brane.
It will now be interesting to study
whether the extra dimensions on the brane in this kind of generalised RS model with higher dimension (e.g. 10d spacetime as required by superstring theory) would provide enough impetus for 3d spcae exponential expansion.
We hope to report these in future works.

This paper is to be published in Chinese Physics C.

\begin{acknowledgments}
We wish to acknowledge the support of the Key Program of National Natural Science Foundation of China (under Grant No. 11535005),
the National Natural Science Foundation of China (under Grant No. 11647087 and No. 11805097),
the Natural Science Foundation of Yangzhou Polytechnic Institute (under Grant No. 201917),
and the Natural Science Foundation of Changzhou Institute of Technology (Grant No. YN1509).
\end{acknowledgments}


\begin{thebibliography}{35}



\bibitem{KK} Th. Kaluza, Sitzungseber. Press. Akad. Wiss. Phys. Math. Klasse 996 (1921); O. Klein, Z. Phys. \textbf{37}, 895 (1926); Nature (London) \textbf{118}, 516 (1926).
\bibitem{NAH1} N. Arkani-Hamed, S. Dimopoulos, and G. Dvali, Phys. Lett. B \textbf{429}, 263 (1998).
\bibitem{NAH2} N. Arkani-Hamed, S. Dimopoulos, and G. Dvali, Phys. Rev. D \textbf{59}, 086004 (1999).
\bibitem{Sundrum} R. Sundrum, Phys. Rev. D \textbf{59}, 085009 (1999).
\bibitem{Lykken} J. Lykken, L. Randall, J. High Energy Phys. \textbf{06}, 014 (2000).
\bibitem{Antoniadis1} I. Antoniadis, Phys. Lett. B \textbf{246}, 377 (1990).
\bibitem{Antoniadis} I. Antoniadis, N. Arkani-Hamed, S. Dimopoulos and G. Dvali, Phys. Lett. \textbf{B 436} 257 (1998).
\bibitem{Polchinski} J. Polchinski, String Theory. Vol. 2: Superstring theory and beyond, Cambridge University Press (1998).
\bibitem{RS} L. Randall and R. Sundrum, Phys. Rev. Lett. \textbf{83}, 3370 (1999).
\bibitem{ADD} N. Arkani-Hamed, S. Dimopoulos and G. Dvali, Phys. Lett. \textbf{B 429} 263 (1998).
\bibitem{Das} S. Das, D. Maity, and S. Sengupta, J. High Energy Phys. 05 (2008) 042.
\bibitem{Koley} R. Koley, J. Mitra, and S. SenGupta, Phys. Rev. D \textbf{92}, 041902(R) (2009).
\bibitem{Burgess} C. P. Burgess, F. Quevedo, G. Tasinato, and I. Zavala, J. High Energy Phys. 11 (2004) 069.
\bibitem{Nilles} H. P. Nilles, A. Papazoglou, and G. Tasinato, Nucl. Phys. \textbf{B677}, 405(2004).
\bibitem{Sasaki} M. Sasaki, T. Shiromizu, and K.-i. Maeda, Phys. Rev. D \textbf{62}, 024008 (2000).
\bibitem{Mitra} J. Mitra, T. Paul, S. SenGupta, Eur. Phys. J. C (2017) 77:833.
\bibitem{SC2} S. Chakraborty, S. SenGupta, Eur. Phys. J. C 75(11), 538 (2015).
\bibitem{Banerjee} I. Banerjee, S. Chakraborty, and S. SenGupta, Phys. Rev. D \textbf{99}, 023515 (2019).
\bibitem{SC1} S. Chakraborty and S. SenGupta, Phys. Rev. D \textbf{92}, 024059 (2015).
\bibitem{SC3} S. Chakraborty and S. SenGupta, Eur. Phys. J. C \textbf{76}, 552 (2016).
\bibitem{Perlmutter} S. Perlmutter et al., Astrophys. J. \textbf{517}, 565 (1999).
\bibitem{Riess} A. Riess et al., Astron. J. \textbf{116}, 1009 (1998).
\bibitem{Bennett} C. L. Bennett et al., Astrophys. J. Suppl. Ser. \textbf{148}, 1(2003).
\bibitem{Netterfield} C. B. Netterfield et al., Astrophys. J. \textbf{571}, 604 (2002).
\bibitem{Halverson} N.W. Halverson et al., Astrophys. J. \textbf{568}, 38 (2002).
\bibitem{Middleton} C. A. Middleton, and E. Stanley, Phys. Rev. D \textbf{84}, 085013 (2011).
\bibitem{Karch} A. Karch and L. Randall, J. High Energy Phys. 05 (2001) 008.
\bibitem{Wu} Z. C. Wu, Phys. Rev. D \textbf{80}, 105001 (2009).




\end{thebibliography}
\end{document}